\documentclass[superscriptaddress,twocolumn,longbibliography,aps,amsmath,amssymb,floatfix,reprint]{revtex4-2}
\usepackage[dvipdfmx]{graphicx}
\usepackage{dcolumn}
\usepackage{bm}
\usepackage[utf8]{inputenc}
\usepackage[T1]{fontenc}
\usepackage{mathptmx}
\usepackage{upgreek}
\usepackage{color}

\begin{document}

\title{Covariance Linkage Assimilation method for Unobserved Data Exploration}

\affiliation{Graduate School of Science and Technology, Nara Institute of Science and Technology, Ikoma, Nara 630-0192, Japan}
\affiliation{Data Science Center, Nara Institute of Science and Technology, Ikoma, Nara 630-0192, Japan}
\affiliation{Research Center for Computational Design of Advanced Functional Materials, National Institute of Advanced Industrial Science and Technology, Tsukuba, Ibaraki 305-8568, Japan}
\affiliation{Center for Computational Sciences, University of Tsukuba, Ibaraki 305-8577, Japan}
\affiliation{Faculty of Science, Tokyo University of Science, Shinjuku, Tokyo 162-8601, Japan.}
\affiliation{Carbon Value Research Center, Research Institute for Science \& Technology, Tokyo University of Science, Noda, Chiba 278-8510, Japan.}
\affiliation{Center for Material Research Platform, Nara Institute of Science and Technology, Ikoma, Nara 630-0192, Japan}

\author{Yosuke Harashima}
\email{harashima.yosuke@ms.naist.jp}
\affiliation{Graduate School of Science and Technology, Nara Institute of Science and Technology, Ikoma, Nara 630-0192, Japan}
\affiliation{Data Science Center, Nara Institute of Science and Technology, Ikoma, Nara 630-0192, Japan}

\author{Takashi Miyake}
\affiliation{Research Center for Computational Design of Advanced Functional Materials, National Institute of Advanced Industrial Science and Technology, Tsukuba, Ibaraki 305-8568, Japan}

\author{Ryuto Baba}
\affiliation{Graduate School of Science and Technology, Nara Institute of Science and Technology, Ikoma, Nara 630-0192, Japan}

\author{Tomoaki Takayama}
\affiliation{Graduate School of Science and Technology, Nara Institute of Science and Technology, Ikoma, Nara 630-0192, Japan}
\affiliation{Data Science Center, Nara Institute of Science and Technology, Ikoma, Nara 630-0192, Japan}

\author{Shogo Takasuka}
\affiliation{Graduate School of Science and Technology, Nara Institute of Science and Technology, Ikoma, Nara 630-0192, Japan}

\author{Yasuteru Shigeta}
\affiliation{Center for Computational Sciences, University of Tsukuba, Ibaraki 305-8577, Japan}

\author{Yuichi Yamaguchi}
\affiliation{Faculty of Science, Tokyo University of Science, Shinjuku, Tokyo 162-8601, Japan.}
\affiliation{Carbon Value Research Center, Research Institute for Science \& Technology, Tokyo University of Science, Noda, Chiba 278-8510, Japan.}

\author{Akihiko Kudo}
\affiliation{Faculty of Science, Tokyo University of Science, Shinjuku, Tokyo 162-8601, Japan.}
\affiliation{Carbon Value Research Center, Research Institute for Science \& Technology, Tokyo University of Science, Noda, Chiba 278-8510, Japan.}

\author{Mikiya Fujii}
\affiliation{Graduate School of Science and Technology, Nara Institute of Science and Technology, Ikoma, Nara 630-0192, Japan}
\affiliation{Data Science Center, Nara Institute of Science and Technology, Ikoma, Nara 630-0192, Japan}
\affiliation{Center for Material Research Platform, Nara Institute of Science and Technology, Ikoma, Nara 630-0192, Japan}

\date{\today}

\begin{abstract}
  This study proposes a materials search method combining a data assimilation technique based on a multivariate Gaussian distribution with Bayesian optimization.
  The efficiency of the search using this method was demonstrated using a pair of example functions.
  By combining Bayesian optimization with the data assimilation technique, 
  the maximum value of the example function was found more efficiently compared to ordinary Bayesian optimization without the data assimilation.
  A practical demonstration was also conducted by constructing a data assimilation model for 
  the bandgap of (Sr$_{1-x_{1}-x_{2}}$La$_{x_{1}}$Na$_{x_{2}}$)(Ti$_{1-x_{1}-x_{2}}$Ga$_{x_{1}}$Ta$_{x_{2}}$)O$_{3}$.
  The concentration dependence of the bandgap was analyzed, 
  and synthesis was performed with chemical compositions in the sparse region of the training data points to validate the predictions.
\end{abstract}

\maketitle

\section{Introduction}

Since the launch of the Materials Genome Initiative~\cite{MGI2011}, 
interest in material development using materials informatics has grown substantially. 
Successful explorations involving millions of compositions have been reported, 
especially focusing on easily collectible data such as formation energy.~\cite{Merchant2023} 
Machine learning models require large datasets for training.
However, in material development, the required properties vary for each material, and the amount of data available is limited. 
For instance, data on properties like the quantum efficiency of photocatalytic compounds, 
magnetocrystalline anisotropy energy of permanent magnet compounds, 
or mechanical properties such as tensile strength or fracture stress, 
are comparatively scarce. 
Such experimental data are difficult to increase in number, 
as their acquisition often involves multiple steps or requires analysis that integrates several raw data.
Therefore, there is a demand for methods that can construct highly accurate predictive models with limited data.

Due to the substantial costs associated with experiments, 
it is impractical to collect a large amount of training data. 
While simulations can provide more data compared to experiments, 
they often contain systematic errors due to approximations, 
rendering them unsuitable for direct prediction of experimental values. 

Data assimilation, which integrates experimental and simulation data, is one of the methods to solve these problems.
Data assimilation has evolved for time-series data such as weather forecasts~\cite{Navon2009,Hamilton2016,Franceschini2020}, 
and recently, attempts have been made to apply it to quantum mechanics~\cite{Giannakis2019}
and materials science.~\cite{Tsujimoto2018,Ohno2020,Yoshikawa2022,Obinata2022,Sakaushi2022} 
A similar concept involves research that integrates two different types of data.~\cite{Fukazawa2019,Iwama2022}
In weather forecasting, 
data assimilation is widely known as a method for combining observational data with numerical model results to make more accurate predictions. 
Similarly, in materials science, 
integrating experimental data with simulation data is expected to provide reliable predictions of material properties and facilitate the discovery of novel materials.

Bayesian optimization has emerged in materials science, 
especially as a transformative tool for the discovery and optimization of new materials.~\cite{Shahriari2016,Garnett2023} 
Evaluations of material properties can be extremely costly and time-consuming, especially for experiments.
Bayesian optimization offers a powerful solution by efficiently navigating the vast, 
often high-dimensional, search spaces for material properties.
This method has been applied to 
mechanical properties,~\cite{Sarvilahti2022} 
elastic properties,~\cite{Balachandran2016}
and crystal structure search.~\cite{Yamashita2018,Sato2020}
Using a probabilistic model, typically a Gaussian process, 
Bayesian optimization develops a predictive understanding of the material behavior based on prior data and uncertainty estimates. 
This approach enables the algorithm to intelligently propose the next best experimental or simulation conditions to rapidly converge on optimal material configurations. 
By balancing the exploration of untested material compositions with the exploitation of promising regions, 
Bayesian optimization accelerates the development of innovative materials with tailored properties. 
This technique is particularly valuable in applications such as designing high-performance alloys, 
engineering advanced ceramics, and optimizing composite materials, 
where each experimental iteration may involve significant resources and time.
We have applied this technique to optimize experimental conditions for polymerization reactions in flow synthesis.~\cite{Takasuka2024} 

In this study, 
we propose a search method that combines a data assimilation technique based on multivariate Gaussian distribution, proposed in a previous study~\cite{Harashima2021} 
with Bayesian optimization to achieve efficient material exploration.
The developed code, CLAUDE (Covariance Linkage Assimilation method for Unobserved Data Exploration), is available here~\cite{CLAUDE2023}.
This method employs a posterior distribution, instead of Gaussian process regression, 
as the strategy for Bayesian optimization.
A key property in the previous study was the treatment of missing data.
Incorporating missing data into the formulation of the posterior distribution 
allowed for the use of independently obtained data in the Bayesian optimization process.
We demonstrated the efficiency of the proposed method with a pair of example functions.
The search for the maximum value of the example function using Bayesian optimization was examined both with and without assimilation.
We also constructed a data assimilation model of the bandgap of (Sr$_{1-x_{1}-x_{2}}$La$_{x_{1}}$Na$_{x_{2}}$)(Ti$_{1-x_{1}-x_{2}}$Ga$_{x_{1}}$Ta$_{x_{2}}$)O$_{3}$ as a practical example of a data assimilation model.
The concentration $(x_{1}, x_{2})$ dependence of the bandgap was analyzed.
Photocatalysts are key materials for addressing energy issues and 
contributing to the realization of a sustainable society.
The bandgap is one of the key factor in photocatalysis for absorbing light.
Subsequently, to validate the predictions, 
the synthesis was performed with chemical compositions in the sparse region of the training data points, 
and the bandgap was observed.

\section{Data assimilation and Bayesian optimization: Formalism}

\subsection{Multivariate Gaussian distribution and prediction model}

Suppose constructing a model to predict a target variable $y := (y_{\mathrm{sim}}, y_{\mathrm{exp}})^{T}$ for a given descriptor $x$, 
where $y_{\mathrm{sim}(\mathrm{exp})}$ denotes a simulational (or experimental) value of target variable.
As in our previous study~\cite{Harashima2021}, 
descriptors $x$ and target variables $y$ are concatenated into a $D$-dimensional vector variable defined as 
$z^{T} := (x^{T},y^{T})$, 
where $z$ is supposed to follow the multivariate Gaussian distribution, 
\begin{equation}
  p(z|\Lambda) = \sqrt{\dfrac{|\Lambda|}{(2\pi)^{D}}} \exp\left(-\dfrac{1}{2} z^{T} \Lambda z\right).
\end{equation}
$\Lambda$ is a precision matrix, which is an inverse covariance matrix.
Here, the components of $\Lambda$ are explicitly expressed for later convenience,
\begin{equation}
  \Lambda = \left(
  \begin{array}{cc}
    \Lambda_{xx} & \Lambda_{xy}
    \\
    \Lambda_{yx} & \Lambda_{yy}
  \end{array}
  \right).
\end{equation}
Given $x$, the conditional probability distribution of the $d$-dimensional target variable $y$ ($d=2$ in this case) is expressed as
\begin{equation}
  p(y|x, \Lambda) = \sqrt{\dfrac{|\Lambda_{yy}|}{(2\pi)^{d}}} \exp\left(-\dfrac{1}{2} (y-\mu)^{T} \Lambda_{yy} (y-\mu)\right),
\end{equation}
where $\mu$ is a mean value of $y$, which is calculated from a given $x$ as
\begin{equation}
  \mu = -(\Lambda_{yy})^{-1}\Lambda_{yx}x.
\end{equation}
Prediction of the target variable is given by this $\mu$ as a function of $x$.
A matrix element of $\Lambda$ for $y_{\mathrm{sim}}$ and $y_{\mathrm{exp}}$ indicates a correlation between simulation and experiment, and 
this correlation assimilates the simulational and experimental data.
A similar example is multitask learning, which is known as a method that leverages the correlations to improve prediction accuracy.~\cite{Breiman1997}

\subsection{Bayesian optimization}

Typically, Bayesian optimization uses Gaussian process regression model, 
however, this study does not use it. 
Instead, we introduce Bayesian optimization using the Monte Carlo method that directly employs the posterior distribution of $\Lambda$ formulated as follows. 
This method generates multiple models with different $\Lambda$s from the posterior distribution. 
Each model provides a different approximation of the objective function, naturally incorporating model uncertainty. 
Through this approach, by aggregating information from multiple models, 
it becomes possible to assess the uncertainty of the prediction model, 
which can be used to propose the next candidate points.

The posterior distribution of $\Lambda$ is written by using Bayes' Theorem as 
\begin{equation}
  p(\Lambda | \{y_{i}\}_{i=1}^{N}, \{x_{i}\}_{i=1}^{N}) = \dfrac{p(\{y_{i}\}_{i=1}^{N}|\Lambda, \{x_{i}\}_{i=1}^{N}) \cdot p(\Lambda|\{x_{i}\}_{i=1}^{N}) }{p(\{y_{i}\}_{i=1}^{N}|\{x_{i}\}_{i=1}^{N})},
  \label{eq:bayestheorem}
\end{equation}
where $p(\Lambda|\{x_{i}\}_{i=1}^{N})$ denotes the prior distribution, which is assumed to be a uniform function in this study.
$N$ is the number of sample data.
The denominator in Eq.~\eqref{eq:bayestheorem} does not depend on $\Lambda$, and thus, 
can be ignored for the distribution of $\Lambda$.
$p(\{y_{i}\}_{i=1}^{N}|\Lambda, \{x_{i}\}_{i=1}^{N})$ is the likelihood.
\begin{align}
  &p(\{y_{i}\}_{i=1}^{N}|\Lambda, \{x_{i}\}_{i=1}^{N}) 
  \nonumber
  \\
  & \qquad = \prod_{i=1}^{N} \sqrt{\dfrac{|\Lambda_{yy}|}{(2\pi)^{d}}} \exp\left(-\dfrac{1}{2} (y_{i}-\mu_{i})^{T} \Lambda_{yy} (y_{i}-\mu_{i})\right),
  \label{eq:likelihood}
  \\
  & \mu_{i} = - (\Lambda_{yy})^{-1} \Lambda_{yx} x_{i}.
  \label{eq:mu}
\end{align}
This is rewritten using matrix $C^{N}$ and a reduced precision matrix $\Lambda^{0}$ as
\begin{equation}
  p(\{y_{i}\}_{i=1}^{N} | \Lambda, \{x_{i}\}_{i=1}^{N}) = \left(\dfrac{\left|\Lambda_{yy}\right|}{(2\pi)^{d}} \right)^{\frac{N}{2}}
  \exp\left(-\dfrac{1}{2} \mathrm{Tr}(C^{N} \Lambda^{0})\right),
  \label{eq:likelihood_covariance}
\end{equation}
where 
\begin{equation}
  C_{\gamma_{1},\gamma_{2}}^{N} := \sum_{i=1}^{N} z_{i, \gamma_{1}} z_{i, \gamma_{2}},
  \label{eq:covariance}
\end{equation}
and 
\begin{align}
  \Lambda^{0} := \left(
  \begin{array}{cc}
    \Lambda_{xy}(\Lambda_{yy})^{-1}\Lambda_{yx} & \Lambda_{xy}
    \\
    \Lambda_{yx} & \Lambda_{yy}
  \end{array}
  \right).
  \label{eq:lambda0}
\end{align}
The derivation of Eq.~\eqref{eq:likelihood_covariance} is described in Appendix~\ref{app:likelihood_covariance}.
The posterior distribution of $\Lambda$ is estimated by using $C^{N}$ which is calculated from a given data set.
By optimizing the posterior distribution in terms of $\Lambda$, 
the Maximum A Posteriori (MAP) probability prediction model is obtained.
Bayesian optimization is performed using this posterior distribution.

\subsection{Missing data and direct likelihood}

\begin{figure}
  \begin{center}
  \includegraphics[width=0.55\linewidth]{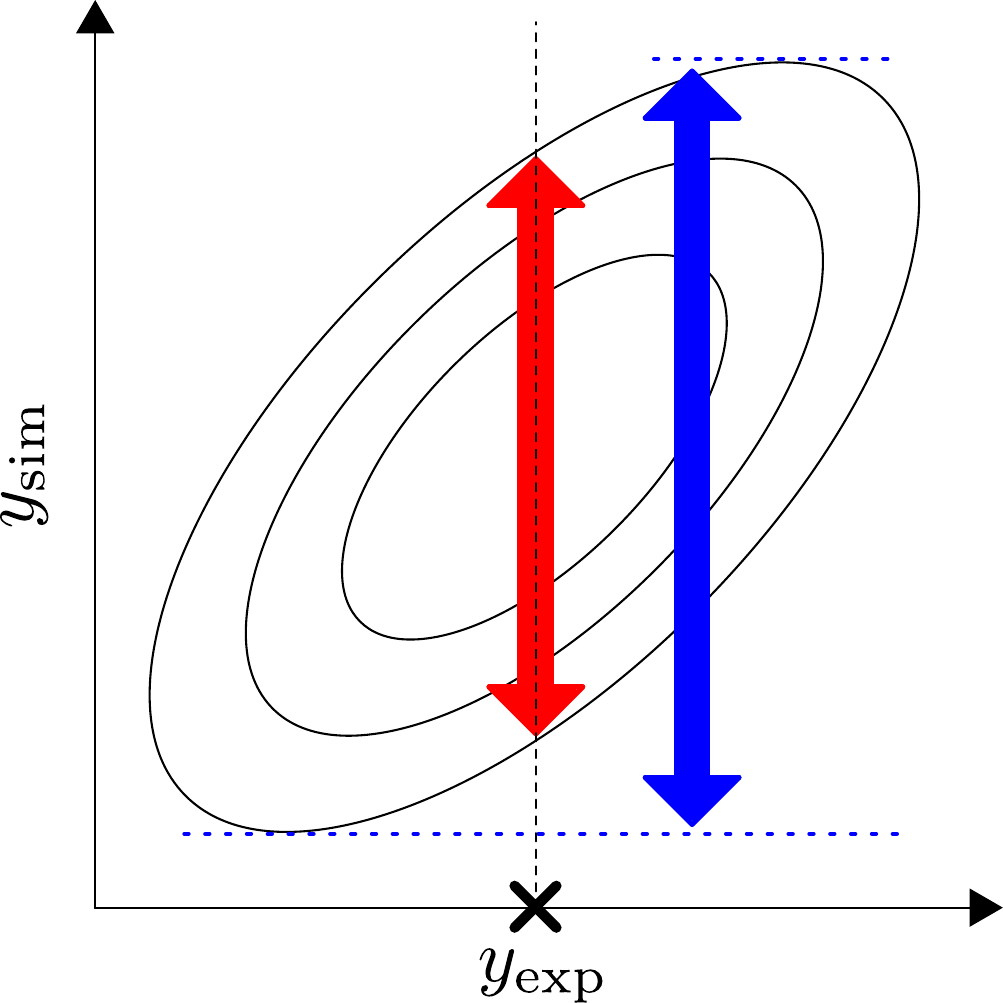}
  \caption{Schematic picture of direct likelihood.
    The cross symbol is a given data of $y_{\mathrm{exp}}$ and 
    the red arrow indicates the distribution width for $y_{\mathrm{sim}}$ with the given $y_{\mathrm{exp}}$.
    The distribution width for $y_{\mathrm{sim}}$ when $y_{\mathrm{exp}}$ is missing is shown as the blue arrow.}
  \label{fig:directlikelihood}
  \end{center}
\end{figure}

Here, we expand the formulation to the missing data. 
The amount of experimental data is usually different from that of simulation data, 
and 
conditions (descriptors $x$) to obtain data are not matched in experiments and simulations.
Those mismatches result in missing values in the experimental or simulation data. 
We use the direct likelihood~\cite{Rubin1976} to address this issue, as in our previous study.~\cite{Harashima2021}
The direct likelihood is defined as the likelihood in which the missing variables are integrated out.
Suppose there are several combinations of missing components, 
for instance, when only simulation data are available, 
the contribution from the missing data is 
\begin{align}
  p(y_{\Gamma}|\Lambda, x) &= \int dy_{\bar{\Gamma}}\; p(y_{\bar{\Gamma}}, y_{\Gamma}|\Lambda, x)
  \nonumber
  \\
  &= \sqrt{\dfrac{|\bar{\Lambda}_{\Gamma\Gamma}|}{(2\pi)^{d_{\Gamma}}}} \exp\left(-\dfrac{1}{2}(y_{\Gamma}-\mu_{\Gamma})^{T}\bar{\Lambda}_{\Gamma\Gamma}(y_{\Gamma}-\mu_{\Gamma})\right),
  \label{eq:directlikelihood}
\end{align}
where $\bar{\Lambda}_{\Gamma\Gamma}$ is Schur complement defined as 
\begin{equation}
  \bar{\Lambda}_{\Gamma\Gamma} := \Lambda_{\Gamma\Gamma} - \Lambda_{\Gamma\bar{\Gamma}} \Lambda_{\bar{\Gamma}\bar{\Gamma}}^{-1} \Lambda_{\bar{\Gamma}\Gamma}, 
\end{equation}
the subscript $\Gamma$ denotes available components, which has $d_{\Gamma}$-dimension ($d_{\Gamma}=1$ in this case), 
and $\bar{\Gamma}$ corresponds missing components.
Figure~\ref{fig:directlikelihood} shows a schematic representation of a distribution of target variables.
When a value of $y_{\bar{\Gamma}}$ is missing, 
the distribution of $y_{\Gamma}$ is broadened through marginalization, 
as indicated by the blue arrow in the figure.
The contribution to the likelihood from the broader distribution decreases, 
and the extent to which the data is reflected in the model diminishes.
The direct likelihood is used instead of the original likelihood in Eqs.~\eqref{eq:likelihood} and \eqref{eq:mu}.
This is equivalent to the replacement of $\Lambda$ with $\bar{\Lambda}$ which is defined as 
\begin{equation}
  \bar{\Lambda} := \left(
  \begin{array}{cc}
    \Lambda_{xx} & \Lambda_{x \Gamma} - \Lambda_{x\bar{\Gamma}}\Lambda_{\bar{\Gamma}\bar{\Gamma}}^{-1}\Lambda_{\bar{\Gamma}\Gamma}
    \\
    \Lambda_{\Gamma x} - \Lambda_{\Gamma\bar{\Gamma}}\Lambda_{\bar{\Gamma}\bar{\Gamma}}^{-1}\Lambda_{\bar{\Gamma} x} & \Lambda_{\Gamma\Gamma} - \Lambda_{\Gamma\bar{\Gamma}} \Lambda_{\bar{\Gamma}\bar{\Gamma}}^{-1} \Lambda_{\bar{\Gamma}\Gamma} 
  \end{array}
  \right).
  \label{eq:lambdabar}
\end{equation}
The derivation is shown in Appendix~\ref{app:lambdabar}.
This $\bar{\Lambda}$ does not have components for missing variables and 
Eq.~\eqref{eq:likelihood_covariance} requires values only for available variables but not for missing ones.
In the $\Lambda^{0}$ calculation of Eq.~\eqref{eq:lambda0}, 
$\bar{\Lambda}$ is used instead of $\Lambda$.
Matrix $C$ in Eq~\eqref{eq:covariance} is calculated separately for each combination of missing components.
Representing the likelihood with matrix $C$ as shown in Eq~\eqref{eq:covariance} 
allows for the sequential updating of the model as new data is obtained, 
thereby removing the necessity to retain previous raw data.

\begin{figure}
  \begin{center}
  \includegraphics[width=\linewidth]{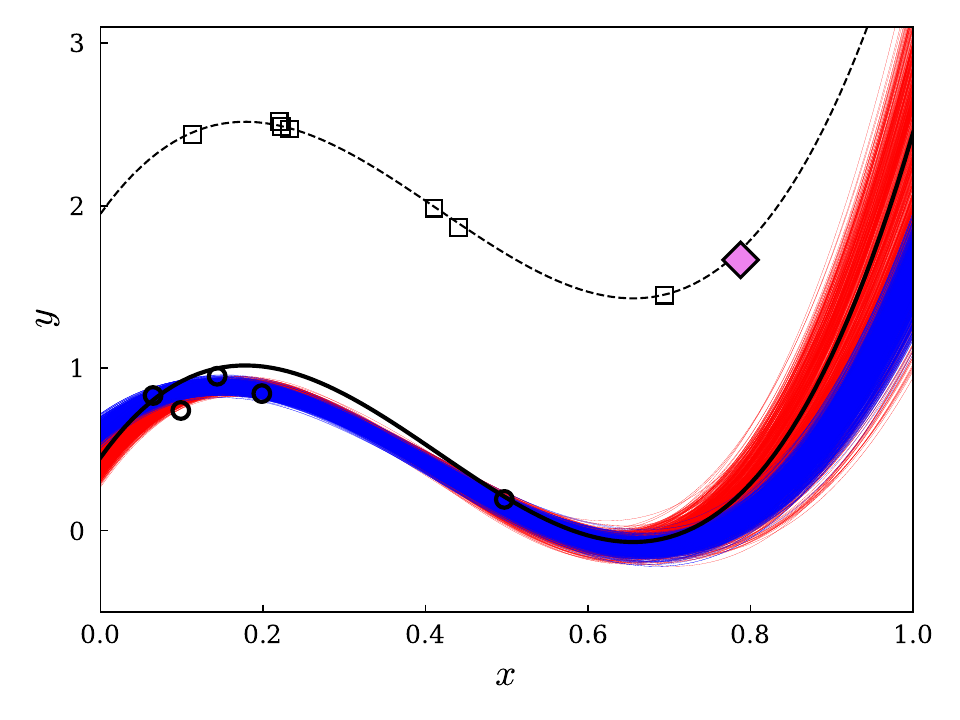}
  \caption{Prediction models generated through Monte Carlo sampling with posterior distribution.
    The same example functions used in our previous study~\cite{Harashima2021} are used.
    The dotted and solid curves are the true model (i) and (ii), respectively.
    Red curves were generated for model (ii), with the posterior distribution trained using the data represented by squares and circles.
    By adding the data for model (i) (represented with diamonds), blue curves for model (ii) were obtained.
    There are 1000 red and blue curves.}
  \label{fig:statistics_update}
  \end{center}
\end{figure}

Figure~\ref{fig:statistics_update} illustrates an effect on one model caused by incorporating data from the other model.
We used the same example functions, (i) and (ii), as those used in our previous work.~\cite{Harashima2021}
\begin{align}
  & f^{\mathrm{i}}(x)=2-x+5(x-0.7)^{2}+20(x-0.5)^{3}, &
  \\
  & f^{\mathrm{ii}}(x)=f^{\mathrm{i}}(x)-1.5. &
\end{align}
The MAP model was obtained using data points represented with squares and circles, 
and 1000 models for model (ii) were generated through Monte Carlo sampling (shown as red curves in the figure).
After adding the new sample data for model (i), the MAP model was retrained, and 
models for (ii) were regenerated based on the updated posterior distribution (shown as blue curves).
It is evident that the distribution of prediction values for model (ii) at $x > 0.7$ decreased after adding the sample data for model (i).
When new data are added to model (i), its distribution becomes narrower. 
Since the distribution of model (ii) is linked to that of model (i), 
the distribution of model (ii) is also narrowed as a result.

\section{Data assimilation and Bayesian optimization: Demonstration}

\begin{figure}
  \begin{center}
  \includegraphics[width=\linewidth]{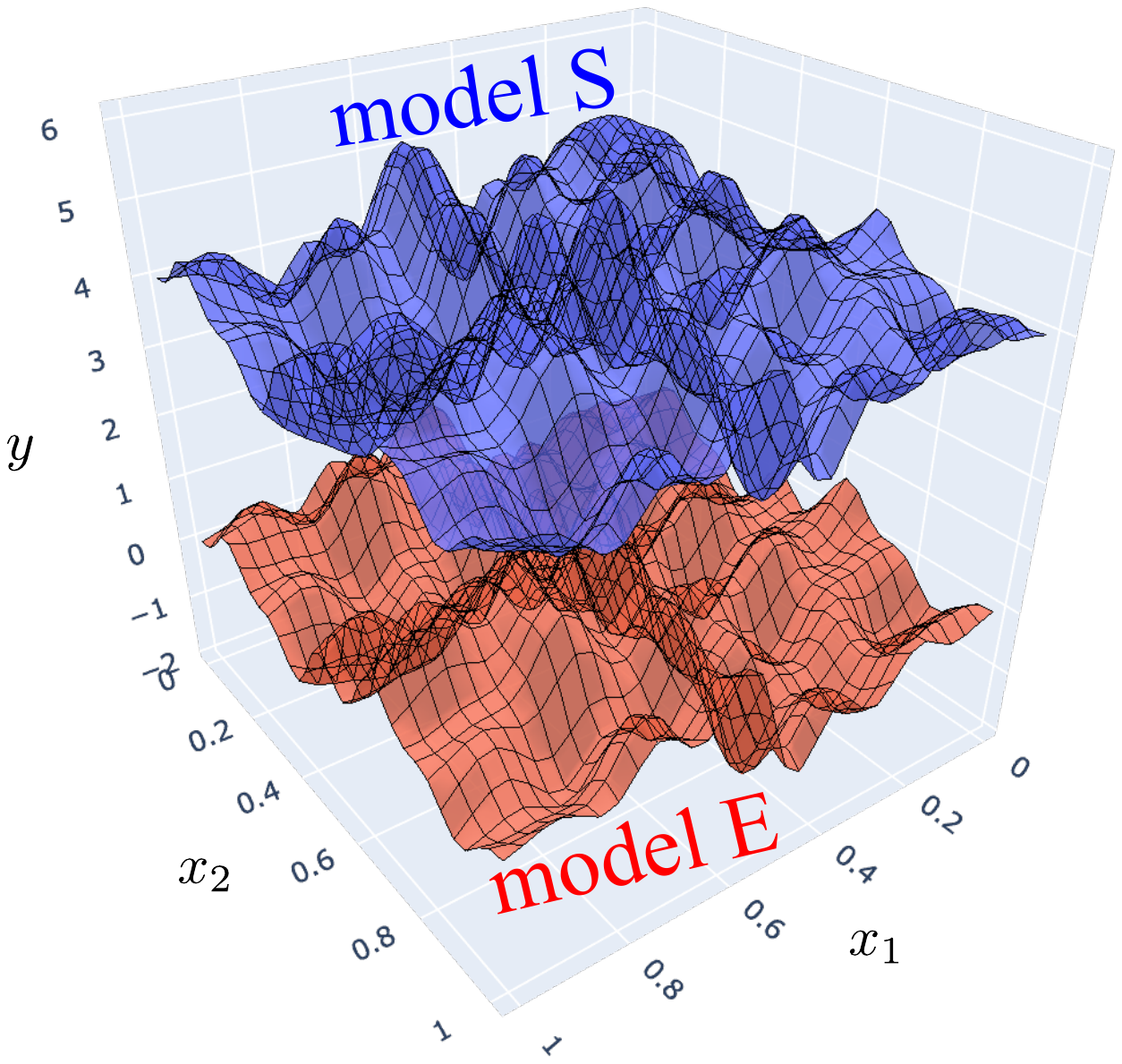}
  \caption{Models E and S are illustrated as red and blue. 
    The models are given as Eq~\eqref{eq:model_demonstration} and Table~\ref{table:coefficient}.
    These models consist of 49 basis set.}
  \label{fig:2d_bo_model}
  \end{center}
\end{figure}

\begin{figure}
  \begin{center}
  \includegraphics[width=\linewidth]{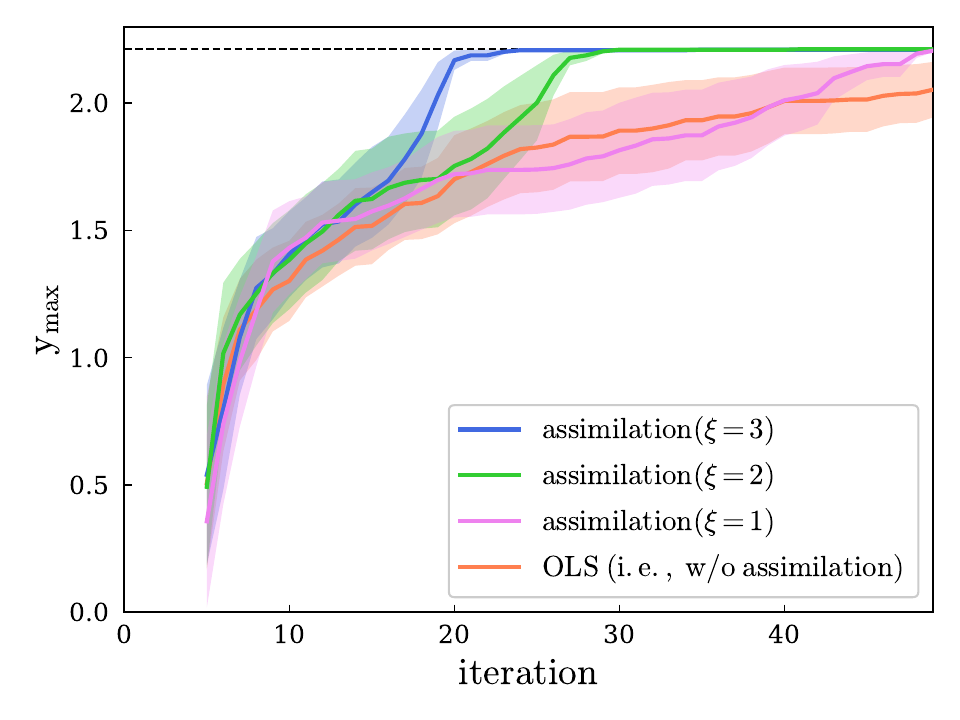}
  \caption{Efficiency of Bayesian optimization using data assimilation. 
    The largest values of the target variable found in the number of samplings are plotted for model E.
    $\xi$ denotes the number of candidates for model S, which varies from 0, 1, 2, and 3, 
    with 0 using the ordinary least square (OLS) model.
    The shaded areas are fluctuations defined in the main body.}
  \label{fig:2d_bo_efficiency}
  \end{center}
\end{figure}

\begin{table*}[t]
  \caption{Coefficients of models S and E. 
    Model E values $w^{\mathrm{E}}_{00}$ and $w^{\mathrm{E}}_{10}$ are shown, 
    values $w^{\mathrm{E}}_{20}$ through $w^{\mathrm{E}}_{35}$ are omitted to avoid overlap with model S.}
  %% \vspace{5pt}
  \begin{center}
    \begin{tabular}{rrrrrrrrrrrrrrrrrr}
      \hline \hline
      \\[-8pt]
      $w^{\mathrm{S}}_{00}$ & $w^{\mathrm{S}}_{10}$ & $w^{\mathrm{S}}_{20}$ & $w^{\mathrm{S}}_{30}$ & $w^{\mathrm{S}}_{40}$ & $w^{\mathrm{S}}_{50}$ & $w^{\mathrm{S}}_{60}$ & $w^{\mathrm{S}}_{01}$ & $w^{\mathrm{S}}_{02}$ & $w^{\mathrm{S}}_{03}$ & $w^{\mathrm{S}}_{04}$ & $w^{\mathrm{S}}_{05}$ & $w^{\mathrm{S}}_{06}$ & $w^{\mathrm{S}}_{11}$ & $w^{\mathrm{S}}_{12}$ & $w^{\mathrm{S}}_{35}$ & $w^{\mathrm{E}}_{00}$ & $w^{\mathrm{E}}_{10}$\\[2pt]
      \hline
      4.0 & 0.2 & 0.3 & $-$0.2 & 0.1 & $-$0.3 & 0.2 & $-$0.1 & 0.4 & 0.1 & $-$0.1 & 0.1 & $-$0.2 & 0.4 & $-$0.2 & $-$0.3 & 0.0 & $-$0.4 \\
      \hline \hline
    \end{tabular}
  \end{center}
  \label{table:coefficient}
\end{table*}

We demonstrated the optimization performance of the aforementioned method for a pair of example models shown in Fig.~\ref{fig:2d_bo_model} as models S and E. 
Both models $f^{\mathrm{S,E}}$ are defined using sine functions as
\begin{equation}
  f^{\mathrm{S,E}}(x_{1},x_{2}) = \sum_{n,m=0}^{6} w^{\mathrm{S,E}}_{nm} \phi_{n}(x_{1})\phi_{m}(x_{2}),
  \label{eq:model_demonstration}
\end{equation}
\begin{equation}
  \phi_{n}(x) := \left\{
  \begin{array}{rc}
    1, & n = 0,
    \\
    \sin(2\pi n x), & n \neq 0.
  \end{array}
  \right.
\end{equation}
The coefficients $w^{\mathrm{S,E}}_{nm}$ are shown in Table~\ref{table:coefficient}.
The degree of freedom of each model is $7 \times 7 = 49$.
We divided the space into a $31 \times 31$ mesh and conducted a search for the maximum value on that mesh.
The maximum value of the model E is 2.21.
Corrections, as discussed in previous work~\cite{Harashima2021}, were included for $(n,m)=(0,0)$ and $(1,0)$ 
to compensate the differences between the models.

Exploration efficiency for the maximum value was examined using a combination of data assimilation and Bayesian optimization.
Workflow of the exploration is as follows.

{\it Step 1.}
Five initial data were sampled randomly for models S and E.

{\it Step 2.}
The prediction model was trained by using sampled data.

{\it Step 3a.}
Candidate values of descriptors were derived from model E.
E stands for experimental data.
These values were set so that an acquisition function was the maximum value.
The acquisition function for model E is chosen as the expected improvement $\alpha_{\mathrm{EI}}$.
\begin{equation}
  \alpha_{\mathrm{EI}}(x) = \dfrac{1}{M}\sum_{k=1}^{M}\max(\mu_{k}(x)-y^{*},0).
\end{equation}
$\mu_{k}$ denotes a predicted value of a model with $k$-th $\Lambda$ given by Monte Carlo sampling according to the posterior distribution.
$y^{*}$ is the current maximum value in the sampled data.
The number of Monte Carlo sampling $M$ was 500.

{\it Step 3b.}
In addition to {\it Step 3a}, some candidates were derived from model S, with S representing simulation data.
We examined three cases: one, two, and three candidates at a time.
The acquisition function for this model was defined as 
\begin{equation}
  \alpha_{\mathrm{Var}}(x) = \dfrac{1}{M}\sum_{k=1}^{M}(\mu_{k}(x)-\mu(x))^{2}.
\end{equation}

{\it Step~2}, {\it 3a}, and {\it 3b} were iterated to find the maximum value.
(For the case of ordinary least square (OLS), {\it Step~3b} was skipped.)
The sets of these steps were repeated 50 times and the performance was analyzed statistically.

Figure~\ref{fig:2d_bo_efficiency} shows the largest values of the target variable found in the number of samplings.
The average values are shown as solid lines in Fig.~\ref{fig:2d_bo_efficiency}, 
and the shaded areas indicate fluctuations $\delta y_{\mathrm{u}}, \delta y_{\mathrm{l}}$ defined as 
\begin{align}
  \delta y_{\mathrm{u}} &:= \dfrac{1}{L}\sum_{l=1}^{L} \max(y^{*}_{l}-\langle y^{*} \rangle,0),
  \\
  \delta y_{\mathrm{l}} &:= \dfrac{1}{L}\sum_{l=1}^{L} \min(y^{*}_{l}-\langle y^{*} \rangle,0),
\end{align}
where $L$ is the number of statistical samplings, which is 50 in this case.
Here, this expression of fluctuations is used instead of the standard deviation, 
which can exceed the upper limit due to a skewed distribution.
The numbers of candidates in {\it Step~3b.}, $\xi$, were 0(OLS), 1, 2, and 3.
The maximum, illustrated as a dashed line, was found by the assimilation more efficiently than the OLS method.
Assuming that data acquisition for model E is costly, while for model S it is inexpensive, 
the search with a larger $\xi$ is more efficient.
The numbers of samplings when the maximum was found were 49, 27, and 20 for $\xi =$ 1, 2, and 3, respectively.
At these numbers, the total number of candidates for model S reached the model degree of freedom, 
at which point the maximum value was suddenly found.
This indicates that 
the assimilation of the established simulation model plays a significant role in identifying the experimental maximum, 
and that, in the process of obtaining simulation data, 
exploration aimed at establishing the simulation model is more effective than exploitation in finding the experimental maximum.

\section{Bandgap of (S\lowercase{r}$_{1-x_{1}-x_{2}}$L\lowercase{a}$_{x_{1}}$N\lowercase{a}$_{x_{2}}$)(T\lowercase{i}$_{1-x_{1}-x_{2}}$G\lowercase{a}$_{x_{1}}$T\lowercase{a}$_{x_{2}}$)O$_{3}$}

\begin{figure}
  \begin{center}
  \includegraphics[width=\linewidth]{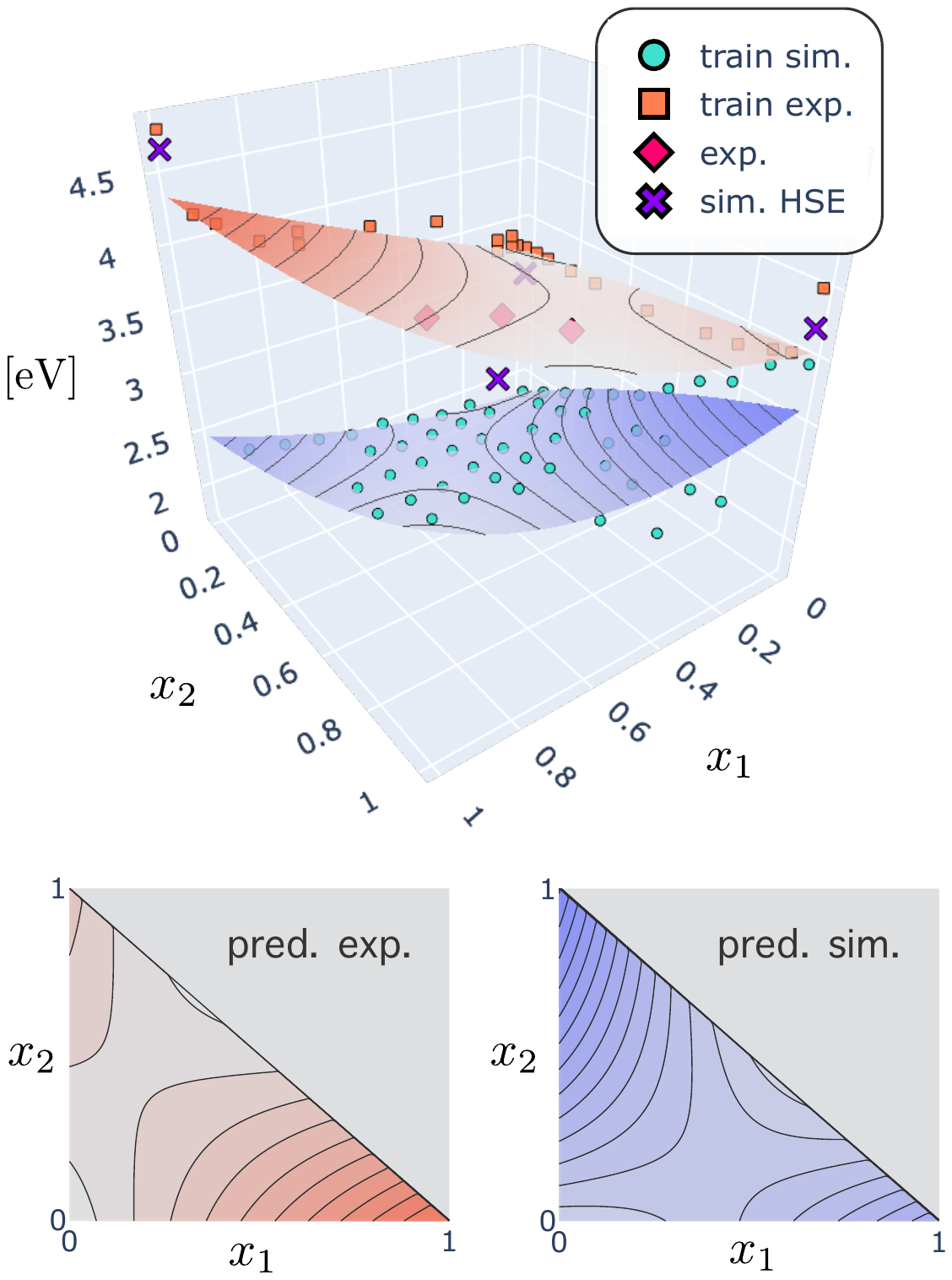}
  \caption{Bandgap of (Sr$_{1-x_{1}-x_{2}}$La$_{x_{1}}$Na$_{x_{2}}$)(Ti$_{1-x_{1}-x_{2}}$Ga$_{x_{1}}$Ta$_{x_{2}}$)O$_{3}$.
    The training experimental data (square) and calculated data (circle) are shown.
    For validating the assimilation model, 
    the test experimental data (diamond) and the HSE calculation data (cross) are also shown.
    The lower two panels are colormaps of the experimental (left) and the simulational (right) prediction functions.}
  \label{fig:bandgap}
  \end{center}
\end{figure}

We also demonstrated the performance of assimilation in a practical case: bandgap values of photocatalytic compounds.
Photocatalysts use the energy of absorbed light for catalytic reactions, e.g., decomposing water to hydrogen and oxygen. 
The wavelength of absorbed light is determined by the bandgap. 
SrTiO$_{3}$, whose crystal structure is perovskite, is a typical photocatalytic compound,~\cite{Tanaka2020,Nishiyama2021}
and formation of solid solution is an effective method to tune bandgaps in photocatalysts.
This is called band engineering and has been extensively studied.~\cite{Osterloh2008,Kudo2009,Wang2020,Yamaguchi2021}

Here, we constructed a model predicting a bandgap for (Sr$_{1-x_{1}-x_{2}}$La$_{x_{1}}$Na$_{x_{2}}$)(Ti$_{1-x_{1}-x_{2}}$Ga$_{x_{1}}$Ta$_{x_{2}}$)O$_{3}$ 
using first-principles calculation and experimental data.
The bandgap values obtained from first-principles calculations are known to qualitatively reproduce experimental values. 
However, the calculated bandgaps tend to show systematic deviations, 
such as being underestimated when using the Perdue-Burke-Ernzerhof generalized gradient approximation~\cite{Perdew1996} (GGA-PBE) for the exchange-correlation energy functional.
These systematic errors are expected to be corrected through assimilation.
The continuous substitution in the calculations was achieved by the Korringa-Kohn-Rostoker (KKR)~\cite{Korringa1947,Kohn1954} Green's function approach and 
the coherent potential approximation.~\cite{Shiba1971,Akai1977}
Calculations were performed for each composition on a uniform mesh; 
however, some did not converge in the self-consistent field calculations and were therefore excluded from the training dataset.
The experimental training data were collected from master's theses written by members of the Kudo group.~\cite{Okutomi2003,Kawahara2009}

The chemical compositions of the training data were 
(Sr$_{1-x_{1}}$La$_{x_{1}}$)(Ti$_{1-x_{1}}$Ga$_{x_{1}}$)O$_{3}$ (solid-state reaction method and polymerized complex method) and 
(Sr$_{1-x_{2}}$Na$_{x_{2}}$)(Ti$_{1-x_{2}}$Ta$_{x_{2}}$)O$_{3}$ (solid-state reaction method) only.
The number of simulation and experimental data points are 54 and 24, respectively.
The root mean squared errors for simulation and experimental training data are 0.13 and 0.17, respectively.
Then, to validate the assimilation model, 
test samples were synthesized in intermediate points away from the training data, 
specifically at $(x_{1},x_{2})=$(0.25,0.5), (0.5,0.25), and (1/3,1/3).
The results are shown in Table~\ref{table:bandgap_test}.
X-ray diffraction analysis showed that the main products were identified to be the perovskite. 
Although little amounts of impurities have been formed, 
their bandgaps are larger than those of the main products 
and therefore do not affect the present results.
Details of the simulations and experiments are provided in Appendices~\ref{app:simulation_bandgap} and \ref{app:experiment_bandgap}, respectively.

\begin{table}[t]
  \caption{Experimental test data for bandgap $E_{\mathrm{g}}$ of (Sr$_{1-x_{1}-x_{2}}$La$_{x_{1}}$Na$_{x_{2}}$)(Ti$_{1-x_{1}-x_{2}}$Ga$_{x_{1}}$Ta$_{x_{2}}$)O$_{3}$.
    The errors between the test data and the predictions are also shown.}
  %% \vspace{5pt}
  \begin{center}
    \begin{tabular}{ccc}
      \hline \hline
      \\[-8pt]
      $(x_{1},x_{2})$ & $E_{\mathrm{g}}$ [eV] & error [eV] \\[2pt]
      \hline
      $(0.25,0.50)$ & $3.32$ & $-0.05$ \\
      $(0.50,0.25)$ & $3.34$ & $-0.22$ \\
      $(1/3,1/3)$   & $3.31$ & $-0.13$ \\
      \hline \hline
    \end{tabular}
  \end{center}
  \label{table:bandgap_test}
\end{table}

\begin{table}[t]
  \caption{Calculated bandgap values of (Sr$_{1-x_{1}-x_{2}}$La$_{x_{1}}$Na$_{x_{2}}$)(Ti$_{1-x_{1}-x_{2}}$Ga$_{x_{1}}$Ta$_{x_{2}}$)O$_{3}$ 
    with the hybrid functional HSE06.~\cite{Krukau2006}}
  %% \vspace{5pt}
  \begin{center}
    \begin{tabular}{ccc}
      \hline \hline
      \\[-8pt]
      $(x_{1},x_{2})$ & $E_{\mathrm{g}}$ [eV] \\[2pt]
      \hline
      $(0,0)$         & $3.02$ \\
      $(1,0)$         & $4.67$ \\
      $(0,1)$         & $3.73$ \\
      $(0.375,0.375)$ & $2.87$ \\
      \hline \hline
    \end{tabular}
  \end{center}
  \label{table:bandgap_hse}
\end{table}

The assimilation model was expanded using a polynomial of the concentrations $x_{1}$ and $x_{2}$,
\begin{equation}
  f^{\mathrm{S,E}}(x_{1},x_{2}) = \sum_{n+m \le 2} w^{\mathrm{S,E}}_{nm} x_{1}^{n} x_{2}^{m}.
\end{equation}
The correction was included for $(n,m)=(0,0)$, $(1,0)$, and $(1,1)$.
Those terms were chosen manually to fit the training data.

The prediction model constructed by assimilating the first-principles and the experimental data is shown in Fig.~\ref{fig:bandgap}.
As previously noted, our calculated values using the GGA-PBE are also found to be underestimated compared to the experimental values,
while the increasing trend along either the $x_{1}$ or $x_{2}$ axis reproduces the experimental data accurately.
It is important to note that the concentration dependence of the bandgap from first-principles calculation is convex downward.
The deviation from the simple linear combination of the bandgap values of LaGaO$_{3}$ and NaTaO$_{3}$ arises from hybridization of electronic states.
The experimental prediction model from data assimilation also retains the convex downward property.
When we look at the actual experimental values for $(x_{1},x_{2})=$(0.25,0.5), (0.5,0.25), and (1/3,1/3), 
it can be seen that they fall below the simple linear combination, 
which can be addressed as an atomic cocktail effect.
Calculated bandgap values with the hybrid functional HSE06~\cite{Krukau2006} were also estimated as shown in Table~\ref{table:bandgap_hse} and Fig.~\ref{fig:bandgap}.
Those bandgap values were 
3.02 eV for SrTiO$_{3}$,
4.67 eV for LaGaO$_{3}$,
3.73 eV for NaTaO$_{3}$, and 
2.87 eV for (Sr$_{0.25}$La$_{0.375}$Na$_{0.375}$)(Ti$_{0.25}$Ga$_{0.375}$Ta$_{0.375}$)O$_{3}$.
The result also shows the downward convexity, 
$ E_{\mathrm{g}} (x_{1},x_{2}) < (1-x_{1}-x_{2}) E_{\mathrm{g}}(0,0) + x_{1} E_{\mathrm{g}}(1,0) + x_{2} E_{\mathrm{g}}(0,1)$ for $(x_{1},x_{2})=(0.375,0.375)$. 
$E_{\mathrm{g}}(x_{1},x_{2})$ denotes a bandgap value for (Sr$_{1-x_{1}-x_{2}}$La$_{x_{1}}$Na$_{x_{2}}$)(Ti$_{1-x_{1}-x_{2}}$Ga$_{x_{1}}$Ta$_{x_{2}}$)O$_{3}$.
Such a convex downward trend in the bandgap induced by multielemental solid solution is an important feature, 
as it enhances visible-light absorption. 
This indicates that our assimilation method can reproduce experimental data.

\section{Conclusion}

In this study, a searching method based on Bayesian optimization combined with data assimilation is proposed.
We have demonstrated two cases: search efficiency and assimilation validity.
We prepared example functions and solved the task of searching for the maximum value.
Compared to the search using the ordinary least squares method, 
the maximum value was found more efficiently using the data assimilation.
To test the validity of the assimilation in a practical case, 
we constructed the assimilation model of the bandgap data for (Sr$_{1-x_{1}-x_{2}}$La$_{x_{1}}$Na$_{x_{2}}$)(Ti$_{1-x_{1}-x_{2}}$Ga$_{x_{1}}$Ta$_{x_{2}}$)O$_{3}$, 
and then synthesized samples for $(x_{1},x_{2})=$(0.25,0.5), (0.5,0.25), and (1/3,1/3).
The bandgap values of the synthesized samples accurately reproduced the downward convex curve of the prediction model.
Although the present study investigated a relatively low-dimensional space, 
even greater effects are expected in high-dimensional spaces---where data tend to be sparse---such as 
the hundreds-dimensional feature spaces generated by featurizers in tools like XenonPy~\cite{Xenonpy2023} or Matminer~\cite{Ward2018}.
The proposed method is not based on Gaussian process regression but rather on a model with explicit basis functions, 
and it is expected that combining it with symbolic regression would enable model selection.
In this study, we considered two types of data---experimental and simulation---but the proposed method can be readily extended to three or more types. 
Its application to transfer learning and multi-fidelity learning involving multiple data sources is also of considerable interest.

\begin{acknowledgments}
The authors sincerely thank Sayo Sugita for her kind assistance in the experiments 
(preparation of a series of test data for (Sr$_{1-x_{1}-x_{2}}$La$_{x_{1}}$Na$_{x_{2}}$)(Ti$_{1-x_{1}-x_{2}}$Ga$_{x_{1}}$Ta$_{x_{2}}$)O$_{3}$ and characterizations of them).
This work was supported by JSPS KAKENHI, Grant-in-Aid for Scientific Research (C) JP22K03449, 
Grant-in-Aid for Scientific Research (A) JP23H00248, 
and MEXT as “Program for Promoting Research on the Supercomputer Fugaku” (realization of innovative light energy conversion materials utilizing the supercomputer Fugaku, Grant Number JPMXP1020210317).
The computation was partly conducted using the facilities of the Supercomputer Center at the Institute for Solid State Physics in the University of Tokyo, 
the Multidisciplinary Cooperative Research Program at the Center for Computational Sciences in University of Tsukuba, 
Subsystem B of the ITO system in Kyushu University, 
and the supercomputer of ACCMS, Kyoto University.
\end{acknowledgments}

\appendix
\section{Derivation of Eq.~\eqref{eq:likelihood_covariance}}
\label{app:likelihood_covariance}

In this section, the derivation of Eq.~\eqref{eq:likelihood_covariance} from Eq.~\eqref{eq:likelihood} is explained.
The product over the samples in Eq.~\eqref{eq:likelihood} becomes the summation on the exponent.
\begin{align}
  &\prod_{i=1}^{N} \sqrt{\dfrac{|\Lambda_{yy}|}{(2\pi)^{d}}} \exp\left(-\dfrac{1}{2} (y_{i}-\mu_{i})^{T} \Lambda_{yy} (y_{i}-\mu_{i})\right)
  \nonumber
  \\
  & \qquad = \left(\sqrt{\dfrac{|\Lambda_{yy}|}{(2\pi)^{d}}}\right)^{N} \exp\left(-\dfrac{1}{2} \sum_{i=1}^{N} (y_{i}-\mu_{i})^{T} \Lambda_{yy} (y_{i}-\mu_{i})\right).
  \label{eq:likelihood_2}
\end{align}
The exponent of Eq.~\eqref{eq:likelihood_2} is expanded to three parts:
$(y_{i})^{T} \Lambda_{yy} y_{i}$, $-(y_{i})^{T} \Lambda_{yy} \mu_{i}$, and $(\mu_{i})^{T} \Lambda_{yy} \mu_{i}$.
Since $\mu_{i}$ can be expressed using $x_{i}$, 
the second term is a cross term between $y_{i}$ and $x_{i}$, 
and the third term is a cross term between $x_{i}$'s.
Consequently, the exponent can be expressed as 
\begin{equation}
  \sum_{i} z_{i}^{T} \Lambda^{0} z_{i} = \mathrm{Tr}(C^{N} \Lambda^{0})
\end{equation}
with a certain $\Lambda^{0}$.
Here, we attempt to derive the actual expression of $\Lambda^{0}$ as follows.

The second term is rewritten by using the explicit formula of $x_{i}$ as 
\begin{align}
  -(y_{i})^{T} \Lambda_{yy} \mu_{i} &= -(y_{i})^{T} \Lambda_{yy} (-(\Lambda_{yy})^{-1}\Lambda_{yx} x_{i})
  \nonumber
  \\
  &= (y_{i})^{T} \Lambda_{yx} x_{i}.
  \label{eq:lambda0_yx}
\end{align}
The third term is 
\begin{align}
  (\mu_{i})^{T} \Lambda_{yy} \mu_{i} &= (-(\Lambda_{yy})^{-1}\Lambda_{yx} x_{i})^{T} \Lambda_{yy} (-(\Lambda_{yy})^{-1}\Lambda_{yx} x_{i})
  \nonumber
  \\
  &= (x_{i})^{T} \Lambda_{xy} (\Lambda_{yy})^{-1} \Lambda_{yx} x_{i}.
  \label{eq:lambda0_xx}
\end{align}
From Eqs.~\eqref{eq:lambda0_yx} and \eqref{eq:lambda0_xx}, 
it is found that $\Lambda^{0}$ is expressed by Eq.~\eqref{eq:lambda0}.

\section{Derivation of Eq.~\eqref{eq:lambdabar}}
\label{app:lambdabar}
Similar to Eq.~\eqref{eq:likelihood_covariance}, with $\bar{\Lambda}$ instead of $\Lambda$,
the exponent of Eq.~\eqref{eq:directlikelihood} is expanded to 
$(y_{\Gamma})^{T} \bar{\Lambda}_{\Gamma\Gamma} y_{\Gamma}$, $-(y_{\Gamma})^{T} \bar{\Lambda}_{\Gamma\Gamma} \mu_{\Gamma}$, and $(\mu_{\Gamma})^{T} \Lambda_{\Gamma\Gamma} \mu_{\Gamma}$.
Then, we derive an explicit form of $\mu_{\Gamma}$.
\begin{equation}
  \left(
  \begin{array}{c}
    \mu_{\Gamma}
    \\
    \mu_{\bar{\Gamma}}
  \end{array}
  \right)
  = - \left(
  \begin{array}{cc}
    \Lambda_{\Gamma\Gamma} & \Lambda_{\Gamma\bar{\Gamma}}
    \\
    \Lambda_{\bar{\Gamma}\Gamma} & \Lambda_{\bar{\Gamma}\bar{\Gamma}}
  \end{array}
  \right)^{-1}
  \left(
  \begin{array}{c}
    \Lambda_{\Gamma x}
    \\
    \Lambda_{\bar{\Gamma} x}
  \end{array}
  \right)
  x.
\end{equation}
The inverse matrix on the right hand side is represented as 
\begin{align}
  &\left(
  \begin{array}{cc}
    \Lambda_{\Gamma\Gamma} & \Lambda_{\Gamma\bar{\Gamma}}
    \\
    \Lambda_{\bar{\Gamma}\Gamma} & \Lambda_{\bar{\Gamma}\bar{\Gamma}}
  \end{array}
  \right)^{-1}
  \nonumber
  \\
  & \qquad = 
  \left(
  \begin{array}{cc}
    \bar{\Lambda}_{\Gamma\Gamma}^{-1} & -\bar{\Lambda}_{\Gamma\Gamma}^{-1}\Lambda_{\Gamma\bar{\Gamma}}\Lambda_{\bar{\Gamma}\bar{\Gamma}}^{-1}
    \\
    -\Lambda_{\bar{\Gamma}\bar{\Gamma}}^{-1}\Lambda_{\Gamma\bar{\Gamma}}\bar{\Lambda}_{\Gamma\Gamma}^{-1} & \Lambda_{\bar{\Gamma}\bar{\Gamma}}^{-1} + \Lambda_{\bar{\Gamma}\bar{\Gamma}}^{-1}\Lambda_{\bar{\Gamma}\Gamma}\bar{\Lambda}_{\Gamma\Gamma}^{-1}\Lambda_{\Gamma\bar{\Gamma}}\Lambda_{\bar{\Gamma}\bar{\Gamma}}^{-1}
  \end{array}
  \right).
  \label{eq:inversematrix}
\end{align}
Using Eq.~\eqref{eq:inversematrix}, $\mu_{\Gamma}$ is written as
\begin{equation}
  \mu_{\Gamma} = -\bar{\Lambda}_{\Gamma\Gamma}^{-1} \left(\Lambda_{\Gamma x} - \Lambda_{\Gamma\bar{\Gamma}}\Lambda_{\bar{\Gamma}\bar{\Gamma}}^{-1}\Lambda_{\bar{\Gamma} x}\right) x.
\end{equation}
Compared to Eq.~\eqref{eq:mu}, it is found that $\Lambda_{yx}$ formally corresponds to $\Lambda_{\Gamma x} - \Lambda_{\Gamma\bar{\Gamma}}\Lambda_{\bar{\Gamma}\bar{\Gamma}}^{-1}\Lambda_{\bar{\Gamma} x}$.
Therefore, $\bar{\Lambda}$ defined in Eq.~\eqref{eq:lambdabar} is formally replaced with $\Lambda$ in Eqs.~\eqref{eq:likelihood} and \eqref{eq:mu}, and
the formally same formulae as Eq.~\eqref{eq:likelihood_covariance}, \eqref{eq:covariance}, and \eqref{eq:lambda0} can be derived for $\bar{\Lambda}$.

\section{Simulation details for analysis of (S\lowercase{r}$_{1-x_{1}-x_{2}}$L\lowercase{a}$_{x_{1}}$N\lowercase{a}$_{x_{2}}$)(T\lowercase{i}$_{1-x_{1}-x_{2}}$G\lowercase{a}$_{x_{1}}$T\lowercase{a}$_{x_{2}}$)O$_{3}$}
\label{app:simulation_bandgap}
Here we show the simulation details for the bandgap analysis of (Sr$_{1-x_{1}-x_{2}}$La$_{x_{1}}$Na$_{x_{2}}$)(Ti$_{1-x_{1}-x_{2}}$Ga$_{x_{1}}$Ta$_{x_{2}}$)O$_{3}$.
The bandgap training data were calculated by using first-principles calculation code AkaiKKR~\cite{Akai2010}
based on Korringa-Kohn-Rostoker~\cite{Korringa1947,Kohn1954} Green's function approach.
The continuous substitution was achieved by coherent potential approximation.~\cite{Shiba1971,Akai1977}
We set $l_{max}=4$ for cation atoms and $=3$ for oxygen atoms.
The parameter \texttt{bzqlty}, which determines the number of k-points, was set to 8.
The lattice constants were estimated from a linear interpolation (Vegard's law) of SrTiO$_{3}$, LaGaO$_{3}$, and NaTaO$_{3}$, 
which were obtained from the structure optimization using the first-principles calculation code, 
the Vienna {\it Ab initio} Simulation Package (VASP)
based on the projector augmented-wave method.~\cite{Bloechl1994,Kresse1996,Kresse1999}
The GGA-PBE~\cite{Perdew1996} was used for the exchange-correlation energy functional.
The energy cutoff was set to 400 eV and $8 \times 8 \times 8$ k-mesh was used.

In order to confirm the convex downward trend of the bandgap with higher accuracy, 
we carried out calculations based on the hybrid functional.
We used the VASP code for the calculations.
The bandgap values of SrTiO$_{3}$, LaGaO$_{3}$, NaTaO$_{3}$, and (Sr$_{0.25}$La$_{0.375}$Na$_{0.375}$)(Ti$_{0.25}$Ga$_{0.375}$Ta$_{0.375}$)O$_{3}$ were calculated.
The unit cells of SrTiO$_{3}$, LaGaO$_{3}$, and NaTaO$_{3}$ were optimized, 
and the lattice constant of (Sr$_{0.25}$La$_{0.375}$Na$_{0.375}$)(Ti$_{0.25}$Ga$_{0.375}$Ta$_{0.375}$)O$_{3}$ was determined by Vegard's law.
A $2 \times 2 \times 2$ supercell was constructed, 
in which the 8 Sr sites were substituted with Sr, La, and Na in a ratio of $2:3:3$. 
The same substitution scheme was applied to the Ti sites.
All distinct substitutional configurations were generated using the supercell code~\cite{Okhotnikov2016}, taking crystal symmetry into account.
A total of 867 symmetry-inequivalent substitutional configurations were generated, 
and the total energy of each was calculated to identify the most stable configuration.
The total energies were calculated with GGA-PBE exchange-correlation functional.~\cite{Perdew1996}
The most stable configuration was chosen from the 867 configurations and 
the bandgap for the structure was estimated from the calculation with the hybrid functional HSE06.~\cite{Krukau2006}

\section{Experimental details for analysis of (S\lowercase{r}$_{1-x_{1}-x_{2}}$L\lowercase{a}$_{x_{1}}$N\lowercase{a}$_{x_{2}}$)(T\lowercase{i}$_{1-x_{1}-x_{2}}$G\lowercase{a}$_{x_{1}}$T\lowercase{a}$_{x_{2}}$)O$_{3}$}
\label{app:experiment_bandgap}
We extracted the details of training experimental data from Okutomi, and Kawahara.~\cite{Okutomi2003,Kawahara2009}
In Fig.~\ref{fig:bandgap}, those data are depicted as squares.
SrTiO$_{3}$-LaGaO$_{3}$ and SrTiO$_{3}$-NaTaO$_{3}$ solid solutions were prepared by a solid-state reaction. 
SrCO$_{3}$(Kanto Chemical: 99.9\%), 
La$_{2}$O$_{3}$(Kanto Chemical: 99.99\%), 
Ta$_{2}$O$_{5}$(Rare metallic: 99.99\%), 
and Na$_{2}$CO$_{3}$(Kanto Chemical: 99.5\%), 
TiO$_{2}$(Soekawa Chemical: 99.9\%), 
Ga$_{2}$O$_{3}$(Kojundo Chemical: 99.99\%) 
were used as starting materials. 
For SrTiO$_{3}$-LaGaO$_{3}$ solid solution, 
the mixture was calcined using an alumina crucible at 1173 K for 1 h. 
The calcined sample was again ground, and then calcined at 1623 K for 10 h to obtain target samples. 
For SrTiO$_{3}$-NaTaO$_{3}$ solid solution, the mixture was calcined at 1423 K for 10-20 h. 
A polymerizable complex method was also employed to obtain a SrTiO$_{3}$-LaGaO$_{3}$ solid solution.
Titanium tetra-n-butoxide (Kanto Chemical: 97.0\%) and citric acid (Aldrich: 99.5\%) were dissolved in ethylene glycol (Kanto Chemical: 99.5\%).
Then, La(NO$_{3}$)$_{3}$-6H$_{2}$O, Ga(NO$_{3}$)$_{3}$-nH$_{2}$O, 
and SrCO$_{3}$ (Kanto Chemical: 99.9\%) were added to the solution and stirred at 363 K for 10-14 h.
The solution was heated at 773 K to promote polymerization between ethylene glycol and citric acid.
The obtained precursor was calcined at 1274-1473 K for 10 h to obtain the solid solution.
The components included in the obtained samples were identified using X-ray diffraction (Rigaku, RINT-1400, Miniflex, Cu K$\alpha$ source). 
Diffuse reflection spectra of the samples were obtained using a UV-vis spectrometer (Jasco; Ubest V-750) 
and were converted from reflection to absorbance by the Kubelka-Munk method.

To verify the validity of the prediction for the bandgaps of (Sr$_{1-x_{1}-x_{2}}$La$_{x_{1}}$Na$_{x_{2}}$)(Ti$_{1-x_{1}-x_{2}}$Ga$_{x_{1}}$Ta$_{x_{2}}$)O$_{3}$,
we actually synthesised them experimentally and measured the bandgap values.
The data is represented as diamonds in Fig.~\ref{fig:bandgap}.
These oxides were prepared by a solid-state reaction.
The starting materials are as follows;
SrCO$_{3}$(99.9\%), La$_{2}$O$_{3}$(99.99\%), Ta$_{2}$O$_{5}$(99.9\%), and Na$_{2}$CO$_{3}$(99.8\%) were purchased from FUJIFILM Wako Pure Chemical;
TiO$_{2}$(99.99\%) was purchased from Kojundo Chemical Laboratory;
Ga$_{2}$O$_{3}$(99.99\%) was purchased from Tokyo Chemical Industry.
The starting materials were mixed by an alumina mortar in a stoichiometric ratio. 
Afterwards, the mixture was calcined using an alumina crucible at 1173 K for 1 h in a muffle furnace.
The calcined sample was ground again, and then calcined at 1423 K for 10 h to obtain target samples.
The components included in the obtained samples were identified using X-ray diffraction (Rigaku, SmartLab, Cu K$\alpha$ source).
Diffuse reflection spectra of the samples were obtained using a UV-vis spectrometer (Jasco; V-750) 
and were converted from reflection to absorbance by the Kubelka-Munk method.

\bibliography{references}

\end{document}